\documentclass[aps,floatfix,onecolumn,superscriptaddress,showpacs,article]{revtex4-1}

\usepackage{float}
\usepackage{amsmath}
\usepackage{amssymb}
\usepackage{graphicx}
\usepackage{bm}
\usepackage{epic}
\usepackage{eepic}
\usepackage{pifont}
\usepackage[utf8]{inputenc}
\usepackage{rotating}
\usepackage{color}
\usepackage{nicefrac}
\usepackage{ulem}
\usepackage[caption=false]{subfig}
\usepackage{array}
\usepackage{tabularx}
\usepackage{soul}
\usepackage{hyperref}
\usepackage{booktabs} 
\usepackage{hyperref}

\begin{document}
\title{Bicuspid Valve Closure and Backflow Prevention:\\
Role of Leaflet Geometry}
\author{B. Kaoui}
\email{badr.kaoui@utc.fr}
\affiliation{
Biomechanics and Bioengineering (UMR 7338),\\
Centre National de la Recherche Scientifique and Universit\'{e} de Technologie de Compi\`{e}gne,\\
60200 Compi\`{e}gne, France}
\author{A. Bou Orm}
\affiliation{
Biomechanics and Bioengineering (UMR 7338),\\
Centre National de la Recherche Scientifique and Universit\'{e} de Technologie de Compi\`{e}gne,\\
60200 Compi\`{e}gne, France}
\author{P. Navet}
\affiliation{
Biomechanics and Bioengineering (UMR 7338),\\
Centre National de la Recherche Scientifique and Universit\'{e} de Technologie de Compi\`{e}gne,\\
60200 Compi\`{e}gne, France}
\author{J. Baish}
\affiliation{
Department of Biomedical Engineering,\\
Bucknell University,\\
Lewisburg, PA 17837, USA
}
\author{L.L. Munn}
\affiliation{
Department of Radiation Oncology,\\
Massachusetts General Hospital and Harvard Medical School,\\
Boston, MA 02114, USA
}
\date{\today}
%
\begin{abstract} 
Bicuspid valves with crescent-shaped leaflets are found in lymphatic vessels and veins, where their primary function is to prevent reflux and ensure unidirectional flow toward the heart. These valves are passive, and their functionality emerges spontaneously from a complex interplay between the properties of the valve leaflets and the flow patterns developing within the vessel sinus region surrounding the valve. 
The main function of the valves is to limit retrograde flow, or reflux, but the optimal valve structure has not been well-characterized. 
Here we investigate numerically how the length of the leaflets affects the valve efficiency in preventing reflux. 
The valves are subjected to backward flow, akin to that imposed by gravity. 
We report the flux through the valve orifice as a function of key parameters: valve length, leaflet length, and leaflet rigidity. 
We monitor the transition in the flow regime - from reflux to complete flow blockage - by varying only the leaflet length. 
The transition threshold is found to depend strongly on the valve shape and stiffness.
We captured these control parameters numerically to evaluate the ability of the valve to close and prevent reflux.
This study allowed us to explain reflux observed experimentally in certain incompetent abnormal and immature valves, particularly those with shorter leaflets.
\end{abstract}
\maketitle
\section{Introduction}
The lymphatic system consists mainly of a ramified network of lymphatic vessels (lymphatics), where lymph fluid is drained from the interstitial tissue by the initial lymphatic capillaries, transported by larger collecting lymphatic vessels, filtered by lymph nodes, and then re-injected back into the thoracic duct and the subclavian veins to reintegrate with the blood circulation \cite{Padera2016,Scallan2016,Moore2018}. 
Other organs, in addition to the vessels, are part of the lymphatic system, such as the lymph nodes, the thymus and spleen.
While the heart serves as the central pump that pushes blood throughout the cardiovascular system, pumping in the lymphatic system is achieved locally at the level of the lymphangion, the vessel segment delimited by two flanking valves.
What is intriguing about this process is the presence of a series of valves along the vessels that guarantee unidirectional flow. 
Each valve opens to allow forward flow, toward the heart, and closes to prevent retrograde flow. 
This occurs spontaneously without any external control \cite{Kunert2015,Munn2015,Baish2016}. 
Biological valves, including lymphatic and venous valves, are passive structures by default. They deform and adapt their configuration in response to the surrounding flow and vessel deformation.
The objective of our present study is to examine the details of the closure mechanisms of bicuspid valves, which are composed of two crescent-shaped leaflets, as found in the lymphatics and blood veins, when subjected to backward flow. 
We aim to extend and complement previous studies that address the open question: 
\textit{What determines the efficiency of bicuspid valve closure during retrograde flow?}
In the present study, we investigate this question with a specific focus: 
\textit{How do leaflet mechanics and geometry affect closure efficiency?}
We address this question for valves of various lengths, and stiffness.
To the best of our knowledge, this aspect is being explored for the first time.
In this work we address an open question in biology and evolution by combining multiphysics simulations with a physics-based approach to analyze fluid–structure interactions and valve mechanics. 
In particular, we investigate a fundamental and still unresolved question: 
\textit{Why has nature produced valve leaflets with a crescent shape?}
Several aspects of valve geometry have already been treated in the literature. 
The effects of valve length and stiffness are relatively well understood: increasing valve length amplifies hydrodynamic resistance, while excessive stiffness hinders proper valve closure. 
However, a crucial question remains largely unexplored and is highly intriguing from both a biological and physical perspective: 
\textit{Why is the leaflet crescent-shaped, and to what extent should the valve leaflet be creased or curved to achieve optimal performance?} In other words, \textit{What is the functional role of the cusp geometry beyond simple valve length and stiffness considerations?}

Most experimental studies on lymphatic vessels are biology oriented. 
Few studies have addressed the challenging \textit{in vivo} and \textit{in vitro} details of the fluid mechanics of the problem, and how fluid forces interplay with the valve mechanics, which is of interest to us. 
The most relevant study is by Munger \textit{et al.}, who investigated the molecular mechanisms underlying lymphatic valve development \cite{Munger2017} and its effect on valve leaflet length, and thus, on the overall valve function. 
They have examined the effect of Connexin43 (Cx43), which is an essential protein for gap junctions between lymphatic endothelial cells. 
The authors used genetically modified mice to assess the consequences of Cx43-specific deletion in the lymphatic system and its role in lymphatic valve development and function. 
They demonstrated that Cx43 deletion in endothelial cells severely impairs valve development, resulting in the overall inability of the valves to prevent reflux. 
As a result, this disruption can give rise to pathologies such as chylothorax. 
On a molecular level, the study identifies Cx43 as a critical component for cell signaling and coordination among lymphatic endothelial cells. 
The results reveal a reduced number of functional valves in mutant mice, exhibiting structural disorganization and defects in morphogenesis, primarily manifesting as valves with short, underdeveloped leaflets that fail to achieve unidirectional flow. 
Another recently published work presents high-quality 3D images of developing lymphatic valves by categorizing valvulogenesis into four distinct stages \cite{Davis2024}. 
Each stage is defined by the growth and extension of the valve leaflets within the vessel. 
At early stages of development, when the leaflet starts taking shape, they are not able to prevent reflux. 
It is not until later stages of development that the mature valve acquires fully developed, long leaflets that can block reflux efficiently.

Computer simulations are used to gain a deeper understanding of the details of valve function and lymphatic vessels, which are not accessible through \textit{in vivo}, or even \textit{in vitro} experiments \cite{Padera2016,Moore2018,Margaris2012}. 
Only a limited number have analyzed valve dynamics systematically using three-dimensional simulations \cite{Watson2017,Wilson2015,Wilson2018,Bertram2020,Bertram2023,Ballard2018,Wolf2021}. 
The most relevant to our study is by Ballard \textit{et al.} \cite{Ballard2018}, who studied the effect of lymphatic valve overall length on its dynamic behavior and function. 
Their simulations focused on the effect of valve length and bending rigidity on valve efficiency. 
The results indicate that, for valves of given elasticity, the overall valve length should be short to optimize forward flow, but long to prevent backflow. 
Indeed, morphological abnormalities or excessive valve rigidity result in inefficient reflux blockage, which is characteristic of pathological conditions. 
The study also highlights mechanisms that may explain valve malfunctions observed in certain pathologies, such as lymphedema, where valves exhibit excessive rigidity or deformability. Although their simulations were also conducted in a straight, rigid vessel, similar to our setup, they did not investigate the role of leaflet length, which is a central focus of the present work. 
The authors developed a model based on a coupled fluid-structure approach. 
Flow dynamics of the lymph were computed using the Lattice Boltzmann Method (LBM) \cite{Succi2001}, while the mechanics of the valves were evaluated with the Lattice Spring Method (LSM) \cite{Ostoja-Starzewski2002}. 
These two numerical methods are also employed in our work, in addition to the Immersed Boundary Method (IBM) for the two-way fluid-structure interaction part \cite{Peskin2002}.

The effect of leaflet length has also been investigated, but in a different context, heart valve implantation. 
Hofferberth \textit{et al.} \cite{Hofferberth2020} proposed an approach conceptually related to our current study on lymphatic valves. 
Their work focuses on artificial cardiac valves and introduces a geometrically adaptable prosthetic valve that can alter its shape in response to pressure and flow variations. Traditional heart valve implants often suffer from anatomical compatibility issues, particularly in patients with complex or atypical geometries. 
Such mismatches can lead to complications, including paravalvular leakage, suboptimal hemodynamic performance, or increased risk of calcification. 
To address these challenges, the authors designed an innovative valve featuring a self-expanding frame and a flexible leaflet structure that together enables patient-specific geometric adaptation. 
Preclinical animal studies demonstrated a substantial reduction in paravalvular leakage owing to the improved anatomical conformity. 
This work builds on an earlier numerical study by Hammer \textit{et al.} \cite{Hammer2017}, who evaluated the performance of bileaflet valves with varying leaflet lengths under progressive vessel dilation. 
Their aim was to model conditions encountered during pediatric growth, where repeated valve replacements are required to match the enlarging cardiovascular anatomy. 
Hammer \textit{et al.} showed that valves with longer leaflets retain the ability to prevent regurgitation even beyond certain degrees of vessel diameter expansion. 
However, their approach did not explicitly account for fluid dynamics or fully coupled fluid–structure interaction (FSI), both of which are incorporated in the present study.

\begin{figure*}[h]
\centering
\includegraphics[width=0.9\textwidth, angle=0]{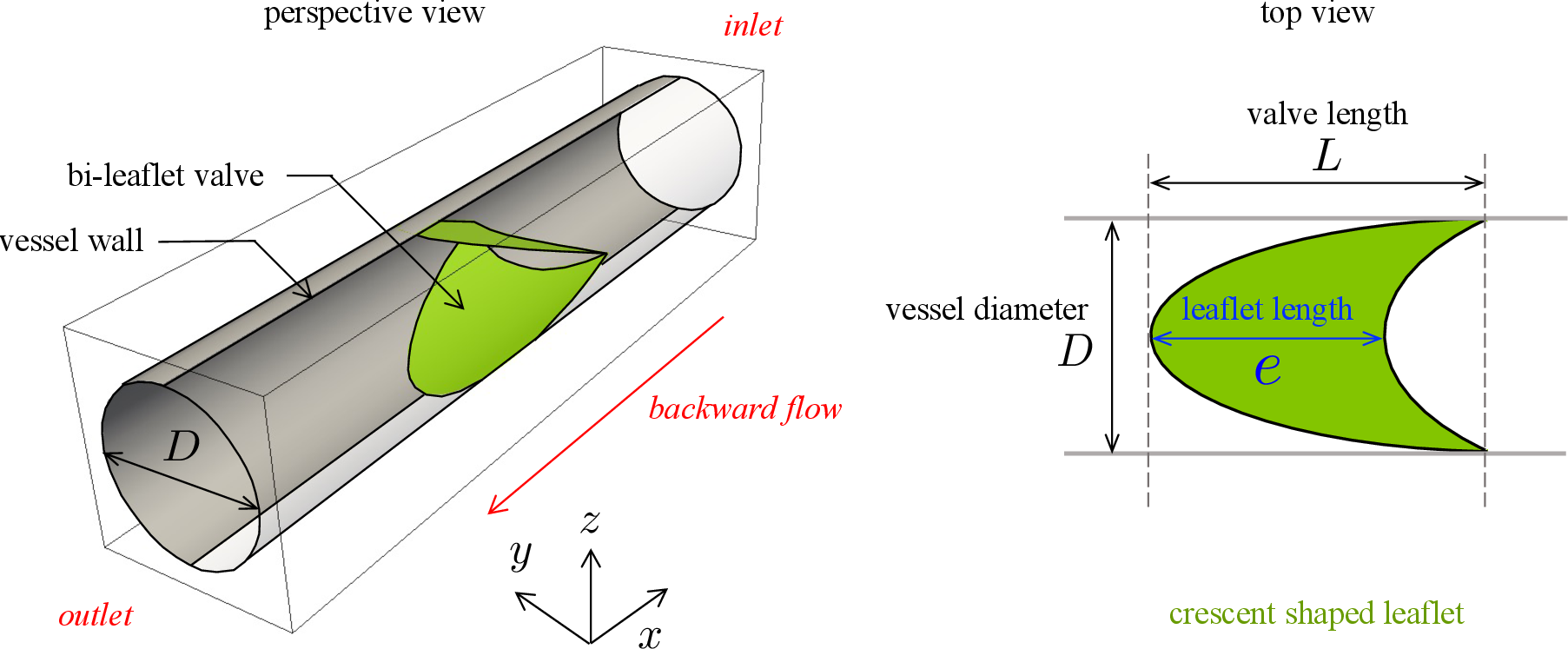}
\caption{
Schematic illustration of the numerical set-up (left panel), consisting of a straight, rigid vessel of circular cross-section enclosing a flexible one-way bi-leaflet valve similar to those found in lymphatic vessels and veins featuring crescent-shaped leaflets (right panel).
}
\label{fig:setup}
\end{figure*}
We complement these previous studies by examining the effect of valve leaflet length, also referred to as cusp length or midline length, on valve performance to close and to prevent reflux. 
We would like to avoid any confusion and clarify that the leaflet length ${\rm e}$ is not the overall valve length ${\rm L}$, see Fig.~\ref{fig:setup}.
In the present study, varying the leaflet length along its centerline corresponds to changing the extent to which the leaflet is scalloped at its center, holding fixed the 
leaflet angle and line of attachment along the vessel wall.
By using leaflet length as a control parameter, we capture the transition from partial reflux to complete flow blockage. 
The critical length at which this transition occurs depends on both valve stiffness and geometric shape parameters. 
In the following sections, we present our computational setup, the methodology, and the resulting flow regimes summarized onto state diagrams. 
Comparisons are made with the available numerical \cite{Hammer2017} and experimental \cite{Munger2017} studies.
\section{Problem statement}
The mechanism of valve closure, induced solely by backward flow, has been the subject of numerous studies in the literature. 
This topic is of particular concern to clinicians due to its association with conditions such as lymphedema in lymphatic vessels and varices in veins. 
From a mechanical perspective, the problem may seem straightforward: the leaflet closes in response to backward flow and subsequently blocks the flow. 
However, the shape, flexibility, and attachment of the leaflet to the vessel wall, as shown in Fig.~\ref{fig:setup}, play a complex and nontrivial role in achieving spontaneous valve closure \cite{Li2019}. 
The ability of the valves to open under forward flow and close under backward flow has been linked to the flow pattern that occurs in the vessel sinus region, which encloses the valve. Some authors have proposed that the primary mechanism of valve closure involves vortices generated within the sinus region, which act to push the two leaflets together to seal and close \cite{Lurie2002,Lurie2003}. 
However, it remains unclear which factors -- valve shape, vessel geometry, or fluid dynamics -- are primarily responsible for the valve's ability to close effectively. 
These questions remain and have been explored by researchers in various disciplines, including biology, fluid mechanics, medicine and biomedical engineering. 
Interestingly, this question has been studied for over five centuries, dating back to Leonardo da Vinci’s early analysis of the flow patterns around heart valves \cite{Wells2013}.

In our present study, we investigate how the shape and elastic properties of the valve leaflets influence the ability of the valve to prevent reflux during imposed backward flow, while maintaining the vessel wall in a straight and rigid configuration (see Fig.~\ref{fig:setup}). 
The flow regime  is given by the dimensionless vessel-based Reynolds number defined as,
\begin{equation} 
{\rm Re}=\frac{DU}{\nu},
\end{equation}
where $D$ is the vessel diameter, $\nu$ is the fluid kinematic viscosity, and $U$ the maximum flow velocity at the vessel axis in the absence of the valve. 
${\rm Re}$ is varied within a small range, less than or equal to unity (${\rm Re} \leq 1$) to model laminar flow in which viscous forces dominate over inertial forces, as found in real lymphatic vessels.
With these assumptions, we aim to isolate two key parameters, the valve elasticity and shape, as the primary contributors, while excluding the effects of inertia, sinus region, and vessel dynamics at this stage.
We report on how the valve geometry, particularly leaflet length (cusp length), and mechanical properties restrict reflux.
\section{Model and Methodology}
We use computer simulations to conduct an extensive parametric study. 
Our simulations are fully three-dimensional and account for the two-way FSI between the valve dynamics and the fluid flow. 
The valve consists of two initially flat crescent-shaped leaflets, as illustrated in Fig.~\ref{fig:setup}. 
They are mathematically assumed to be of zero thickness. 
To model the effect of gravity in a limb when a person is in a vertical standing posture, the flow is applied in the backward direction, opposite to $x$-direction, by exerting a constant, uniform body force on the fluid. 
The valve has a projected length ${\rm L}$, while each leaflet has a projected length or elongation ${\rm e}$, measured at its midline. 
${\rm L}$ and ${\rm e}$ are not the actual lengths, but rather the projections on the x-y plane, as shown in Fig.~\ref{fig:setup}. 
In this framework, the absence of a valve corresponds to ${\rm e}=0$, whereas a fully extended valve, with a flat edge line, corresponds to ${\rm e}={\rm L}$. 
The leaflet's free-moving edge is defined by a parabola. 
The leaflet rigidity $\kappa$ in our study is uniform and arises from the resistance of the valve to three deformation modes \cite{Provot1995}: stretching-compression, shear and bending, with rigidities set to be of the same value to ensure that none of the three deformation modes becomes artificially dominant in our simulations.
They therefore contribute with comparable intensities and to reduce the problem's mechanical parameters to a single parameter.
We use the associated stiffness dimensionless number,
\begin{equation}
{\rm K} = \frac{\kappa}{\eta U},
\end{equation} 
obtained by making the Navier-Stokes equations, with the forcing term, dimensionless (see Appendix A).
$\eta$ is the dynamic viscosity of the fluid. 
The parameter ${\rm K}$ controls the valve's overall rigidity. 
Higher values correspond to stiffer valve leaflets, and lower values to softer ones. 
We approach the problem from a fluid mechanics perspective, disregarding biochemical signaling \cite{Kunert2015,Baish2016,Li2019}, and focusing solely on the resulting flow patterns and their interaction with the valve.

The simulations rely on a combination of three complementary numerical methods  \cite{Kaoui2025}: 
i - the Lattice Boltzmann Method (LBM) used to compute the fluid dynamics \cite{Succi2001}. 
LBM is particularly well-suited for predicting flow in complex geometries, such as those found in porous media, microfluidic devices, and lymphatic and cardiovascular vessel networks. 
It is also compatible with parallel computing due to its localized nature, allowing for faster simulations; 
ii - the Lattice Spring Method (LSM) employed to recapitulate the elasticity of valve leaflets \cite{Ostoja-Starzewski2002,Provot1995}. 
This approach enables the modeling of deformable structures in a simple manner, without resorting to a more sophisticated combination of continuum solid mechanics and finite element methods; 
iii - the Immersed Boundary Method (IBM) used to achieve the two-way coupling of FSIs \cite{Peskin2002}. 
It represents effectively how lymph flow deforms the valve leaflets and, in turn, how these deformations perturb the surrounding flow. 
The computer code developed for this study is openly accessible via a GitLab repository \cite{gitlab}.
\section{Results and Discussion}
\subsection{Valve Closure Performance}
\begin{figure*}
\centering
\includegraphics[width=1\textwidth, angle=0]{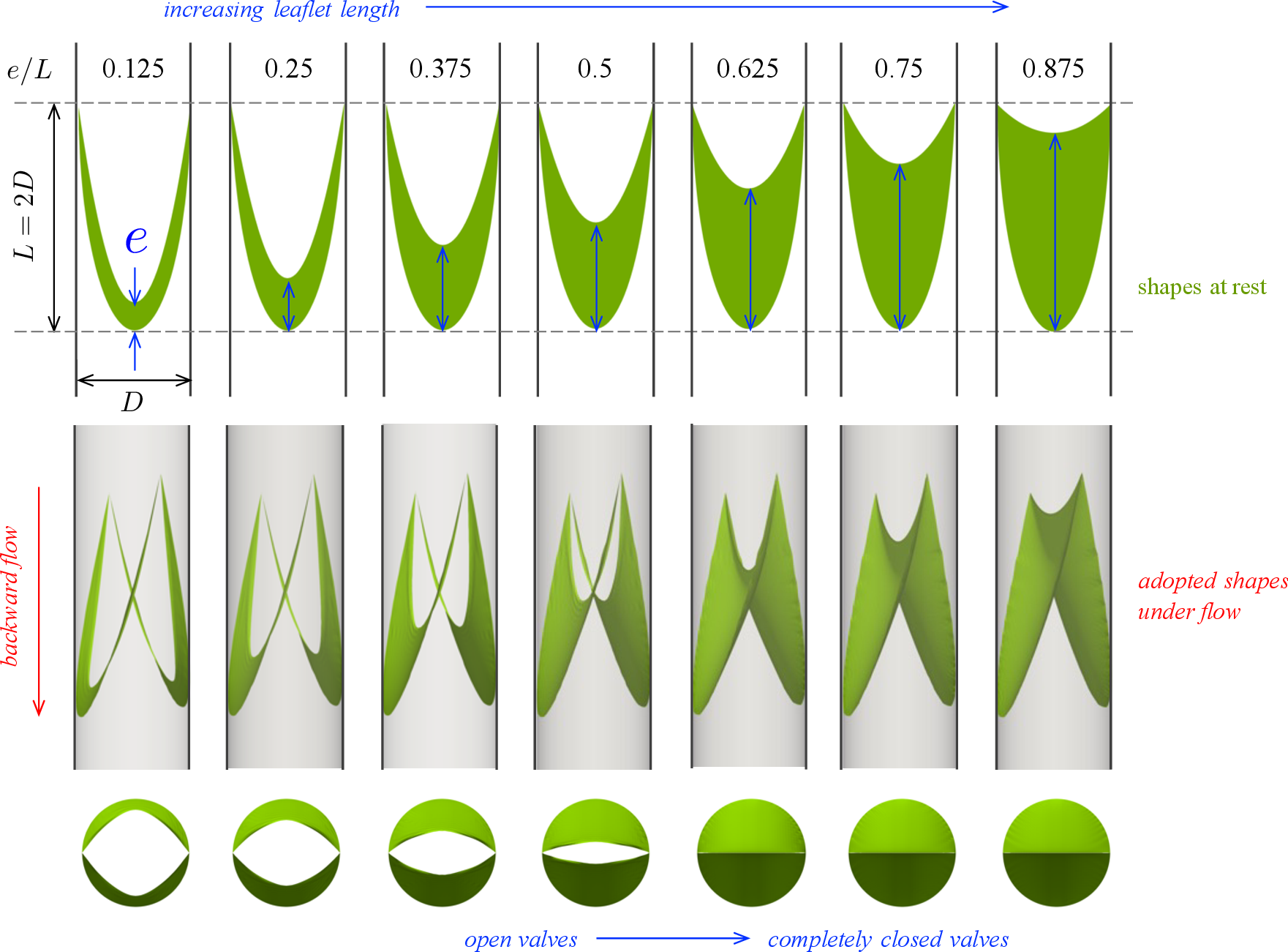}
\caption{
(top panel) Top view of the resting shape of valves of the same length (${\rm L} = 2{\rm D}$), but with varying leaflet midline lengths ${\rm e}$. (middle panel) Semi-profile view of the corresponding deformed shapes in the steady regime under backward flow, slightly rotated to better display leaflet deformation. (bottom panel) Frontal view showing that valves with shorter leaflets form a wider open orifice, which fails to prevent reflux, whereas valves with longer leaflets close completely and thus successfully block reflux.
}
\label{fig:shapes}
\end{figure*}
The initial and equilibrium shape of the leaflets is assumed to be flat and crescent-shaped. The applied flow is steady and driven by a uniform body force. 
All the results shown in the following figures are obtained at Reynolds number ${\rm Re}=1$.
After a transient numerical phase, the valve reaches a steady configuration with a shape that depends on the elastic and geometric properties of the leaflets.  
Figure~\ref{fig:shapes} shows the steady-state shapes adopted by valves of the same length (${\rm L}=2{\rm D}$), but with varying leaflet lengths ${\rm e}$. 
Valves with longer leaflets develop a curved closed shape when exposed to backward flow. 
In contrast, valves with shorter leaflets fail to effectively block lymph flow, allowing fluid leakage through their wide open orifice. 
To ensure the formation of a coaptation zone, and thus complete valve closure, special contact treatment was necessary for valves with soft, longer leaflets, see Appendix B. 
Shorter leaflets lack sufficient surface area to obstruct the flow effectively, while medium-length leaflets have just enough surface to significantly reduce the flow rate. 
In contrast, valves with longer leaflets have excess surface area beyond the necessary minimum. 
The region enclosed by the valve leaflet surface and the vessel wall acts as a pocket, capable of receiving fluid during backward flow. 
The softer the valve, the more it deforms as the fluid impinges on it. 
As a result, the two leaflets move closer together and form a coaptation zone, ensuring complete closure of the valve. 
This configuration effectively blocks backward flow and eliminates reflux.
\subsection{Reflux Quantification}
We quantify valve efficiency in preventing reflux by measuring the net flow rate ${\rm Q}$ across a cross section of the vessel away from the valve location,
\begin{equation}
{\rm Q} = \iint_A u(y, z) \, {\rm d}A,
\end{equation}
where $u(y,z)$ is the local fluid velocity, along the vessel axis, in the $x$-direction, at a point $(y,z)$ belonging to the cross-sectional plane defined by $y$ and $z$, ${\rm d}A$ is the infinitesimal area element in the $y-z$ plane, and $A$ the cross-sectional area of the vessel.
\begin{figure*}[t]
\centering
\includegraphics[width=0.5\textwidth, angle=0]{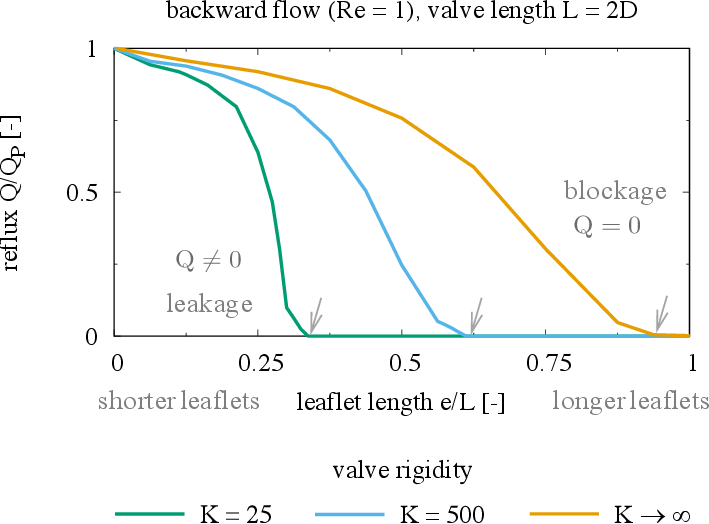}
\caption{
Reflux ${\rm Q}$ as a function of leaflet length ${\rm e}$ for various valve rigidities ${\rm K}$, as defined in the main text.
No interpolation is used. 
The data points are connected with straight lines, using a finer step size in the vicinity of the transition point. Flexible valves successfully prevent reflux beyond a certain leaflet length (indicated with arrows), shifting toward shorter leaflets for softer valves. 
The orange curve represents completely rigid valves with nondeformable and flat leaflets, which fail to close and thus prevent reflux.
}
\label{fig:rigidity}
\end{figure*}

The flow rate is computed as a function of various key parameters. 
For example, we analyze how the flow rate varies with the valve leaflet length ${\rm e}$ for valves of different rigidities ${\rm K}$ (see Fig.~\ref{fig:rigidity}). 
The leaflet length is scaled by the valve overall length ${\rm L}$, yielding a ratio that ranges from $0$ to $1$. 
In the absence of a valve (${\rm e}=0$), the flow rate corresponds to pure Poiseuille flow in a tube of length $l$, which can be theoretically estimated using the formula, 
\begin{equation}
{\rm Q}_{\rm P} = \frac{\pi R^4}{8 \eta} \left( \frac{\Delta p}{l} \right)= \left( \frac{\pi \nu D}{8} \right) {\rm Re},
\end{equation}
where $\Delta p = p_{\rm in} – p_{\rm out}$ is the applied pressure gradient, where $p_{\rm in}$ and $p_{\rm out}$ are, respectively, the pressures set at the inlet and the outlet of the vessel of radius $R=D/2$.
When a valve is present (${\rm e} \neq 0$), the flow rate decreases as the valve obstructs the flow, eventually reaching zero when the central part of the leaflets is  sufficiently long. The flow rate is expressed as a ratio of the computed flow rate to the Poiseuille flow rate ${\rm Q}/{\rm Q}_{\rm P}$. 
As the leaflet length increases, the flow rate gradually declines monotonically to zero. The transition from a leaking valve (${\rm Q} \neq 0$) to complete flow blockage (${\rm Q}=0$) occurs at different thresholds (indicated with arrows) depending on the valve stiffness. 
More flexible valves can block the flow even with shorter leaflets, as their flexibility enables a more complete and hermetic closure. 
From this study, we conclude that, for valves of the same length and stiffness, those with longer leaflets are more effective at preventing reflux than those with shorter leaflets.

\begin{figure*}
\centering
\includegraphics[width=1.\textwidth, trim=0 0 0 0, clip]{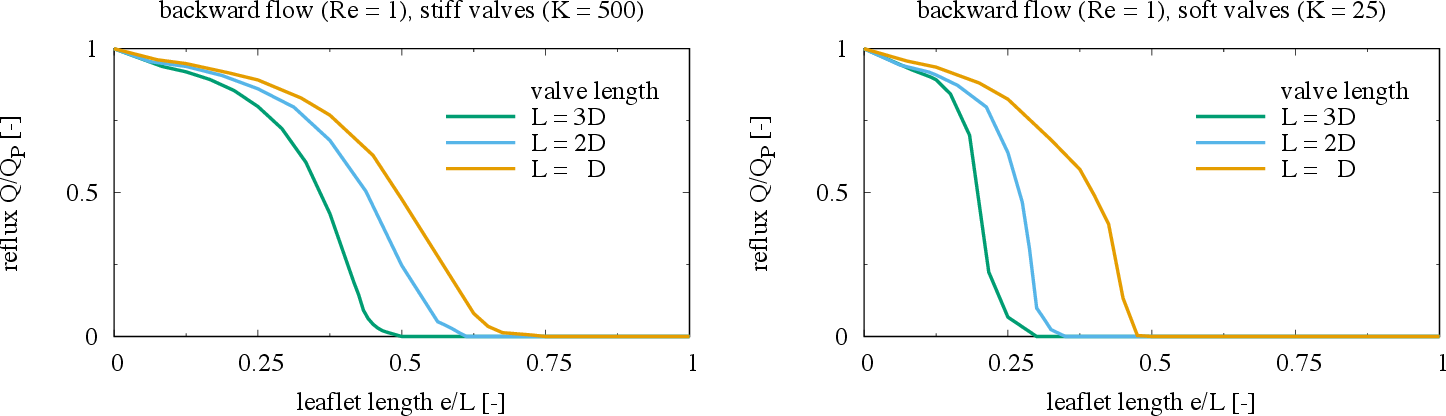}
\caption{Effect of the valve length ${\rm L}$ on valve performance in preventing reflux, particularly on the shift of the threshold that determines the transition from a leaky (${\rm Q} \neq 0$) to a sealed state (${\rm Q}=0$). The left panel shows the results for stiff valves, and the right panel for soft valves. No interpolation is used. The data points are connected with straight lines, using a finer step size in the vicinity of the transition point.
Longer and more flexible valves prevent reflux effectively, even though their leaflets are shorter.}
\label{fig:comp}
\end{figure*}

We also examined the ability of the valves to block the flow, while varying the valve overall length ${\rm L}$. 
We recomputed the flow rate ${\rm Q}$ as a function of leaflet length ${\rm e}$ for different valve lengths ${\rm L}$, considering both stiff and soft valves, with ${\rm K} = 500$ and ${\rm K} = 25$, respectively. 
Figure~\ref{fig:comp} shows a reduction in the reflux with increasing leaflet length. 
Additionally, longer valves are expected to further reduce flow within the vessel because they occupy a larger portion of it, thereby increasing hydrodynamic resistance \cite{Ballard2018}. 
This holds true regardless of the valve stiffness.
The critical leaflet length required to completely block the flow decreases as the valve length increases. 
This effect is more pronounced for soft valves and less significant for stiff valves. 
Based on this study, we conclude that longer soft valves are effective at preventing reflux, even though they have shorter leaflets.
\subsection{Competency-state diagram}
\begin{figure*}[b]
\centering
\includegraphics[width=1.\textwidth, angle=0]{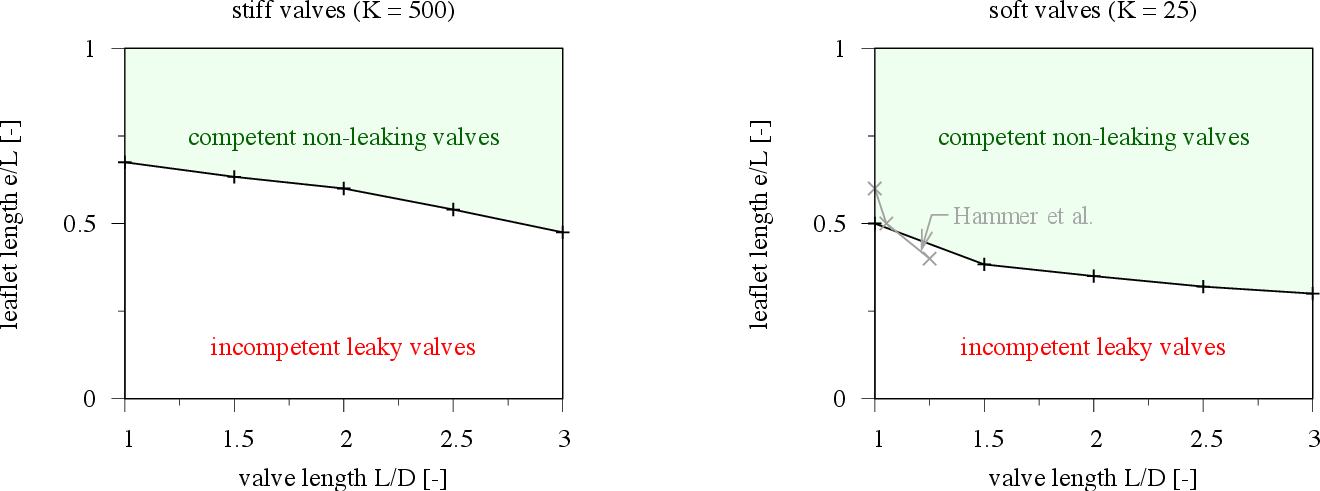}
\caption{State diagrams summarizing the performance of one-way bicuspid valves, with crescent-shaped leaflets, in preventing reflux during backward flow. 
The left panel shows the results for stiff valves (${\rm K}=500$), and the right panel shows the results for soft valves (${\rm K}=25$). 
Incompetent valves are leaky, allowing fluid to pass through their adopted open orifice, whereas competent valves close completely, preventing reflux. 
The transition boundary shifts upward for stiffer valves, while the competence region for softer valves spans a broader area of the diagram. The gray curve in the right panel corresponds to simulations from Ref.~\cite{Hammer2017} that give the borderline above which the valves are competent. They show good agreement with our results.
}
\label{fig:diagram}
\end{figure*}
We summarize our results in a state-diagram representation (Fig.~\ref{fig:diagram}), where we plot the critical leaflet length as a function of the valve length ${\rm L}$. 
The axes are scaled by ${\rm D}$ (horizontal) and ${\rm L}$ (vertical). 
The left panel of Fig.~\ref{fig:diagram} shows the threshold in the limit of stiff valves, and the right panel corresponds to the softest valves that can be modeled numerically.
The region below the boundary denotes incompetent, leaky valves, which fail to prevent reflux because they maintain an open orifice under backward flow. 
In contrast, the region above the boundary corresponds to competent nonleaking valves that completely close and block reflux, even when a counteracting flow-driven body force is maintained.
These diagrams allow classification of a valve as competent or incompetent based solely on its geometric parameters. 
They are particularly useful for elucidating how geometry governs the ability of biological valves to prevent reflux.

\newpage
We compare our results with those obtained for bicuspid valve implants by Hammer \textit{et al.} \cite{Hammer2017}. 
In this work, the effect of vessel radius growth has been studied, with the initial valve configuration completely sealed and featuring nonflat, crescent-shaped leaflets. 
The simulations were used to examine whether the valve leaflets remain sealed as the vessel diameter increases, thereby mimicking growth in children. 
Incompetent valves develop an interleaflet orifice that allows the undesired reflux. 
We specifically report the critical vessel diameter (scaled by valve length) identified in Ref.~\cite{Hammer2017} for a given leaflet length, beyond which the valves remain sealed. 
The right panel of Fig.~\ref{fig:diagram} shows a gray curve connecting three data points from  Ref.~\cite{Hammer2017} mapped onto our proposed state diagram representation. 
The simulations of Ref.~\cite{Hammer2017} were done in the absence of explicit direct interaction with the ambient fluid effects. 
The valve deformation results from applying a prescribed pressure field, rather than from a two-way FSI. 
However, the results still show good agreement with our data obtained for the soft valves.
This comparison reveals the same trend observed in our study: the critical leaflet length decreases as the overall valve length increases. 
\subsection{Comparison with experimental observations}
\begin{figure*}[b]
\centering
\includegraphics[width=0.5\textwidth, angle=0]{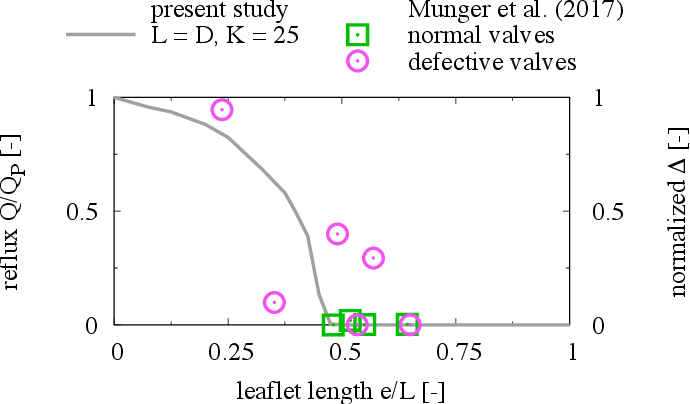}
\caption{Experimental data points ($\Delta$ on the right $y$ axis) extracted from Ref.~\cite{Munger2017} compare the performance of normal and defective lymphatic valves, which were treated with protein during their development to alter their leaflet length development and their ability to prevent reflux. The few experimental data points match well with our simulation curve (${\rm Q}$ on the left $y$-axis) computed for short (${\rm L}={\rm D}$) and soft (${\rm K}=25$) valves with varying leaflet lengths. The transition from leaky (${\rm Q}\neq 0$, $\Delta \neq 0$) to sealed valves (${\rm Q} = \Delta = 0$) occurs around ${\rm e}/{\rm L}=0.5$, as we predict numerically.}
\label{fig:munger}
\end{figure*}
The ranges of real lymphatic valve overall length ${\rm L}$ and leaflet length ${\rm e}$ are not well documented in the literature.
For the valve length ${\rm L}$, we refer to the only available experimental measurements in rats, which show that the ratio of valve length to vessel diameter, ${\rm L}/{\rm D}$, varies between $1$ and $3$ \cite{Ballard2018}.
The leaflet length ${\rm e}$ has been measured only in Refs.~\cite{Sabine2012,Munger2017,Davis2024}. 
The study in Ref.~\cite{Munger2017} demonstrated that leaflet length can be modified during valve morphogenesis through protein treatments. 
To the best of our knowledge, this is the only experimental study that systematically measures leaflet length and investigates its impact on the valve's ability to prevent reflux. 
The discrete points in Fig.~\ref{fig:munger} represent the experimental quantity $\Delta$, and the values are reported on the right $y$ axis.
$\Delta$ quantifies the deviation of the upstream (inlet) pressure, measured during leakage, from the control pressure, measured in the absence of leakage, while the downstream (outlet) pressure is held constant (see the annotations and definitions used in Ref.~\cite{Munger2017}). 
The $x$ axis reporting ${\rm e}$ is scaled using ${\rm L} = 74.5\,\mu{\rm m}$ for normal valves and ${\rm L} = 59.3\,\mu{\rm m}$ for abnormal valves.
$\Delta$ is normalized to allow direct comparison with our numerical results, shown by the continuous curve in the same plot. 
We chose ${\rm L} = {\rm D}$ to match the experimental setup, and ${\rm K} = 25$ to represent softer valves, consistent with the conditions in the experiments of Ref.~\cite{Munger2017}. 
Thus, both the geometry and the stiffness in the simulations were selected to correspond to realistic experimental parameters. 
For normal valves, which possess sufficiently long leaflets to block the imposed reflux, $\Delta = 0$.
Nonzero values of $\Delta$ correspond to abnormal valves with altered leaflet growth. 
In these cases, reduced leaflet length decreases the hydrodynamic resistance of the valve, leading to leakage and an amplification of backward flow, as reported in Ref.~\cite{Munger2017}. 
The shorter the leaflets relative to the overall valve length, the larger the value of $\Delta$.
The good qualitative agreement between the experimental observations and our numerical predictions provides a clear physical explanation of the data.
Our aim here is not to validate the numerical method using the experiments, but rather to help predict and interpret the experimentally obtained results, which are limited in number and exhibit significant scatter.
Our results capture clearly the transition in flow regime - from reflux (${\rm Q} \neq 0$) to complete flow blockage (${\rm Q} = 0$) - induced solely by varying the leaflet length ${\rm e}$ as a control parameter.
Most importantly, the transition occurs near ${\rm e}/{\rm L} = 0.5$ for both the numerical and experimental parameter sets, revealing the existence of a precise critical leaflet length beyond which the valve fully suppresses backward flow.
These findings are also consistent with the conclusions of Davis \textit{et al.} \cite{Davis2024}: immature valves with short leaflets at early stages of development are leaky, whereas mature valves with fully developed long leaflets are competent and effectively prevent reflux.
\section{Conclusions}
We present the first analysis of the performance of one-way bicuspid valves, made of two crescent-shaped leaflets, in preventing lymphatic reflux under gravity-driven flow.
By extending the leaflet midline length as a control parameter, we capture the transition from limited reflux to complete flow blockage, with the critical leaflet length found to depend on both valve overall length and leaflet stiffness.
Through a systematic parametric study, we identify the combination of geometric and mechanical parameters that determine the efficiency of valve closure, and we summarize these results in a state-diagram representation. 
Our findings span a broad parameter space and show good agreement with the limited numerical and experimental data available in the literature. 
We expect that the boundary location in the state diagram will be influenced by additional factors, such as the anisotropy and the nonhomogeneity of leaflet stiffness, the vessel sinus region, and vessel wall flexibility, which warrant further investigation.

Although the typical leaflet length is not well-documented, most observed lymphatic valves exhibit a crescent shape and show various extensions depending on their stage of development and mechanical properties. 
In our study, the flow was maintained within the creeping, laminar flow regime, typical of lymphatic vessels, where inertial effects are negligible \cite{Moore2018}. 
The findings are also expected to apply to venous valves, which share similar geometric features with lymphatic valves, characterized by bicuspid valves with two crescent-shaped leaflets enveloped in a sinus region of veins, but operating at higher Reynolds numbers and controlling the flow of blood.
\section*{Acknowledgment}
BK, JB, and LLM thank Huabing Li for valuable discussions. 
BK also acknowledges fruitful exchanges with Abdul Barakat and Jocelyn Etienne. 
BK, ABO, and PN acknowledge Marc Villegas for technical support to run simulations on the PILCAM2 supercomputing facilities of the Université de Technologie de Compiègne.
BK received financial support from the BMBI laboratory and the French National Research Agency under grant ANR-20-CE45-0008.
\section{Data availability}
The data that support the findings of this article were generated using computer code developed specifically for this study, which is openly accessible via a GitLab repository \cite{gitlab}.
\section*{Appendix A: Nondimensionalization of the NS eqs}
The Navier-Stokes equations (NS) governing the flow of an incompressible Newtonian
fluid is given by
\begin{equation}
\rho\left(
\frac{\partial \mathbf{u}(\mathbf{x},t)}{\partial t}
+ \mathbf{u}(\mathbf{x},t)\cdot\nabla \mathbf{u}(\mathbf{x},t)
\right)
= - \nabla p(\mathbf{x},t) + \eta \nabla^{2}\mathbf{u}(\mathbf{x},t)
+ \mathbf{g}(\mathbf{x}) + \mathbf{f}(\mathbf{r},t)\delta(\mathbf{x} - \mathbf{r}),
\label{eq:NS}
\end{equation}
where $\rho$ and $\eta$ are the mass density and viscosity of the fluid, in this study, the lymph. 
The solutions are obtained for the velocity field $\mathbf{u}$ and the pressure field $p$ at time $t$. 
$\mathbf{g}$ is a constant uniform body force directed opposite to the $x$ axis (see Fig.~\ref{fig:setup}), used to induce backward flow. 
$\mathbf{f}$ is the force exerted by a point belonging to the valve $\mathbf{r}$ upon the fluid point $\mathbf{x}$. 
It has a nonzero value only at $\mathbf{r}$, and zero elsewhere. 
We introduce the following dimensionless variables,
\[
\mathbf{x}^* = \frac{\mathbf{x}}{D},\qquad
t^* = \frac{t}{T},\qquad
\mathbf{u}^* = \frac{\mathbf{u}}{U},\qquad
p^* = \frac{D}{\eta U}p,
\]
using the vessel diameter $D$ as a characteristic length, the flow velocity at the vessel centerline in the absence of the valve $U$, the timescale $T=D/U$ as a characteristic time.
Substituting these variables into Eq.~\eqref{eq:NS} and multiplying
by $D^2/(\eta U)$ yields the dimensionless NS equations without the forcing terms,
\begin{equation}
\mathrm{Re}\, \left[  \frac{\partial \mathbf{u}^*}{\partial t^*}
+ \mathbf{u}^*\cdot\nabla^*\mathbf{u}^* \right]
= -\nabla^* p^* + \nabla^{*2}\mathbf{u}^*,
\label{eq:scaled2}
\end{equation}
where $\mathrm{Re}=\frac{DU}{\nu}$ is the Reynolds number.

Next we express the valve force in dimensionless form. 
We use the Lattice Spring Method (LSM)~\cite{Ostoja-Starzewski2002,Provot1995}, which
models the valve structure as a network of points interconnected by springs.
Each point $i$ in the network experiences a force from its neighboring points $j$, governed by Hooke's law,
\begin{equation}
\mathbf{f}(\mathbf{r},t) = \mathbf{f}(i) = 
-\kappa \sum _{j \neq i}
[ d(i,j) - d_0(i,j) ] 
\frac{\mathbf{d}(i,j)}{\| \mathbf{d}(i,j) \|},
\label{eq:hook}
\end{equation}
where $\kappa$ is the spring rigidity, $d$ and $d_0$ are the current and initial equilibrium distances between points $i$ and $j$, respectively, and $\mathbf{d}$ the position vector connecting these points. 
Multiplying Eq.~\eqref{eq:hook} by $\delta(\mathbf{x}-\mathbf{r})
= \frac{1}{D^3}\delta(\mathbf{x}^*-\mathbf{r}^*)$ and by $D^2/(\eta U)$ yields,
\begin{align}
\frac{D^2}{\eta U}\mathbf{f}(\mathbf{r})\delta(\mathbf{x}-\mathbf{r})
=&\
-\frac{\kappa}{\eta U} \sum _{j \neq i}
[ d^*(i,j) - d_0^*(i,j) ] 
\frac{\mathbf{d}^*(i,j)}{\| \mathbf{d}^*(i,j) \|}
\delta(\mathbf{x}^* - \mathbf{r}^*).
\end{align}
This introduces $\mathrm{\rm K} = \frac{\kappa}{\eta U}$, the valve stiffness dimensionless number.

\section*{Appendix B: Methods for Simulating Closed Blocking Valves}
The numerical method developed for the present study is an extension of our previously well-established models and methods, now adapted to the specific case of the valve dynamics. 
In earlier works, we applied these methods to both 2D and 3D deformable structures, including flowing soft particles (see, e.g., Refs. \cite{Kaoui2022,Krueger2014,Dupin2007}). 
Modeling valve dynamics presents additional numerical challenges, due to the presence of a freely moving part, the valve leaflets, with curved and inclined edges anchored to the vessel wall, see Fig.~\ref{fig:setup}. 
These challenges are particularly pronounced when using the Immersed Boundary Method (IBM) to accomplish the Fluid-Structure Interaction (FSI). 
Moreover, modeling how deformable and anchored valve leaflets can fully close and completely halt a continuously forced fluid flow is a highly demanding computational task. 
The technical details of the numerical methods are described in detail elsewhere \cite{Kaoui2025}. 
Here, we only focus on highlighting the key features that enabled us to achieve full valve closure and complete blockage of the backward flow driven by a steady body force. 
In the absence of a correct solution, we cannot quantitatively assess the numerical error of our approach.
Results need to be compared against a reference that is not known a priori and, to the best of our knowledge, has not been established in the literature. 
We can therefore only evaluate the expected results only qualitatively rather than quantitatively.
The following three subsections present each encountered numerical issue and its solution.

\begin{figure*}[h]
\centering
\includegraphics[width=0.5\textwidth, angle=0]{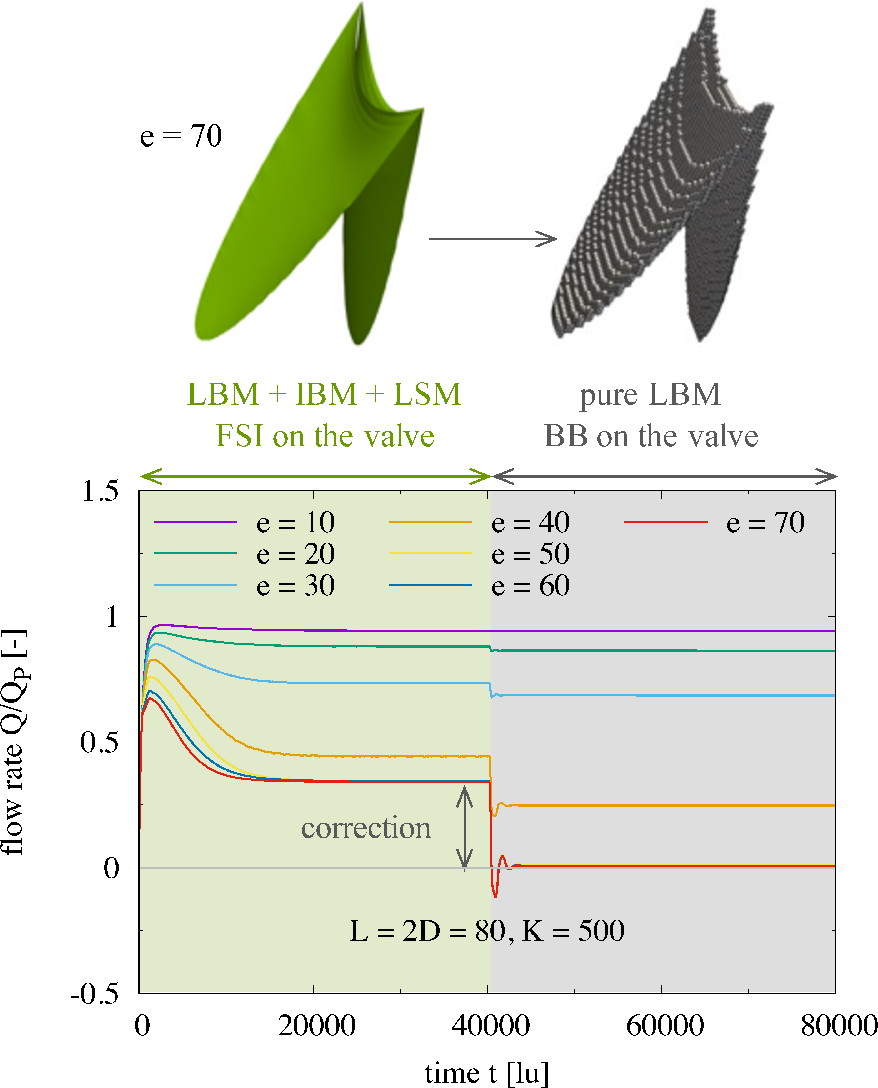}
\caption{
Apparent fluid leakage through the leaflet, arising from the Immersed Boundary Method (IBM), is suppressed by applying bounce-back boundary condition (BB) on each valve leaflet once the solution reaches steady state. The reflux over time, both before and after applying the correction, is compared in the lower panel. Leakage is more prominent for longer leaflets, and the correction effectively suppresses this unphysical leakage.
}
\label{fig:bb_bc}
\end{figure*}
\subsection{Fluid leakage through the leaflets}
We unexpectedly measured  a non-zero reflux, despite the valve leaflets appearing fully closed in all simulations. In these simulations, we use the Immersed Boundary Method (IBM) to handle the Fluid–Structure Interaction (FSI), and the bounce-back condition (BB) of the Lattice Boltzmann Method (LBM) to set a non-slip boundary condition on the vessel wall. A deviation of approximately $25\%$ from the expected zero reflux is observed, see the evolution over time of the flow rate at the early stages of the simulations in the lower panel of Fig.~\ref{fig:bb_bc}. 
For valves with longer leaflets, the reflux reaches up to $25\%$ of the imposed flow rate ${\rm Q}_{\rm P}$, whereas we should ideally achieve zero reflux. We suspect this numerical issue arises from an apparent fluid leakage through the valve leaflets, which are intended to be perfectly impermeable.

The leaflets are modeled as a network of points interconnected by springs in the Lattice Spring Model (LSM), which has been shown to exhibit leakage in the present study. 
However, in principle, the IBM should enforce the impermeability of the leaflet. 
To the best of our knowledge, we have correctly implemented the IBM scheme for the two-way valve–fluid interaction. 
Similar IBM implementations have successfully enforced impermeable boundary conditions for immersed structures, as confirmed by analyzing streamline behavior near the boundary  (see Ref. \cite{Kaoui2011}), the streamlines do not cross the boundaries. 
In those cases, the structures were free-moving soft particles, whereas in the present case, the key difference is the presence of anchored boundaries, where the valve leaflets are attached to the vessel wall. 
This anchoring configuration is implemented by enforcing zero net force and freezing the displacement at the attachment points.

To eliminate artificial reflux, we apply the LBM BB condition not only to the vessel wall, but also to the valve leaflets when the valve reaches its steady shape. 
Initially, the simulation runs with full FSI and BB applied solely on the vessel wall. 
After the valves stabilize, FSI is turned off, and BB is applied on both the vessel wall and the valve leaflets, see the upper panel of Fig.~\ref{fig:bb_bc}. 
This correction successfully eliminates the reflux, achieving the expected zero net flow for valves with longer leaflets. 
In contrast, for valves with shorter leaflets, the deviation from zero flow rate is negligible. This correction was possible because the system tends toward a steady state at longer times. However, we believe it remains important to further investigate the root cause of IBM failure to fully enforce leaflet impermeability in this setup. This will be an important direction for future work.
\subsection{Partial closure of valves with longer leaflets}
\begin{figure*}[h]
\centering
\includegraphics[width=0.5\textwidth, angle=0]{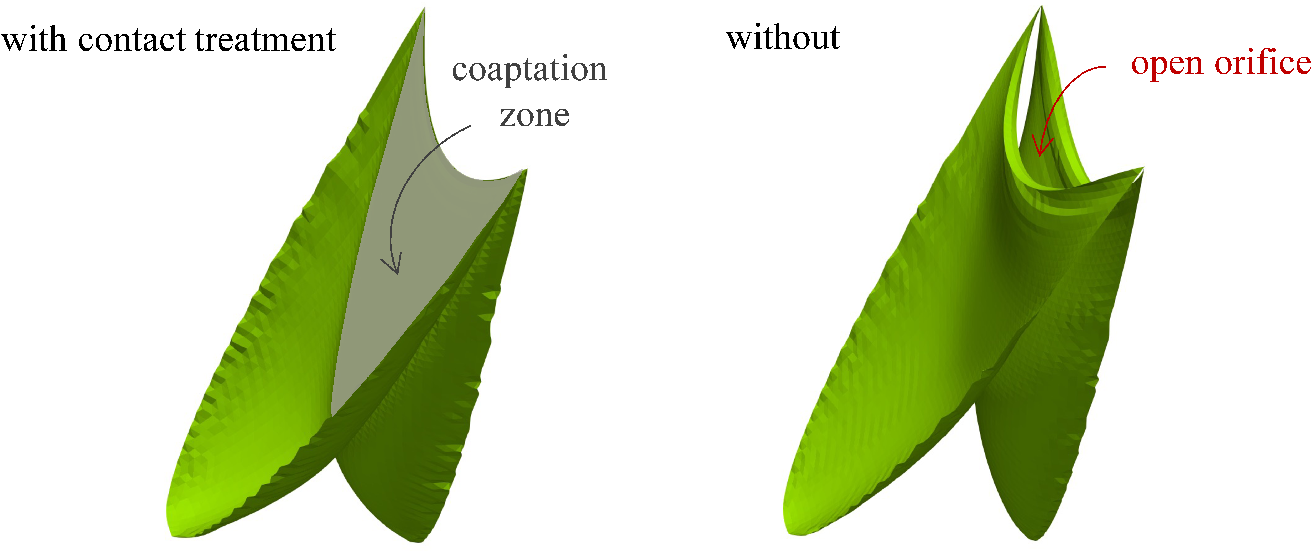}
\caption{
Comparison between the steady-state shapes of soft, longer valves with longer leaflets: the left panel shows results with contact treatment, leading to a flat coaptation zone, and the right panel shows results without contact treatment, characterized by incomplete valve closure.}
\label{fig:coaptation}
\end{figure*}

Simulations of soft valves with longer leaflets show incomplete closure, as seen in the right panel of Fig.~\ref{fig:coaptation}. 
The valves remain partially open, with an open orifice gap forming due to the influence of hydrodynamic forces at low Reynolds numbers, which push the leaflet free edges away from the vessel centerline.

To enhance valve closure, a short-range attractive force is applied between the free-moving parts of the two leaflets. 
This force draws the leaflets closer together until the interleaflet distance becomes sufficiently small, at which point a counteracting repulsive force prevents numerical attachment or overlapping. 
This interaction promotes a more realistic and complete closure, characterized by the formation of a flat coaptation zone, see the right panel of Fig.~\ref{fig:coaptation}, similar to those observed \textit{in vivo} for lymphatic and venous valves, as well as in some valve computer simulations that utilize contact treatment.

\newpage
The implemented force varies exponentially with distance, with its intensity decaying away from the vessel center $x$-$y$ plane. 
The force magnitude must be carefully tuned according to the leaflet stiffness. 
While we are not aware of venous or lymphatic valves being inherently sticky, it is worth noting that most biological tissues are coated with biochemical substances that may give rise to weak and short-range attractive interactions.
\subsection{Non-zero net flux during valve closure}
\begin{figure}[h]
\centering
{\includegraphics[width=0.45\textwidth, angle=0]{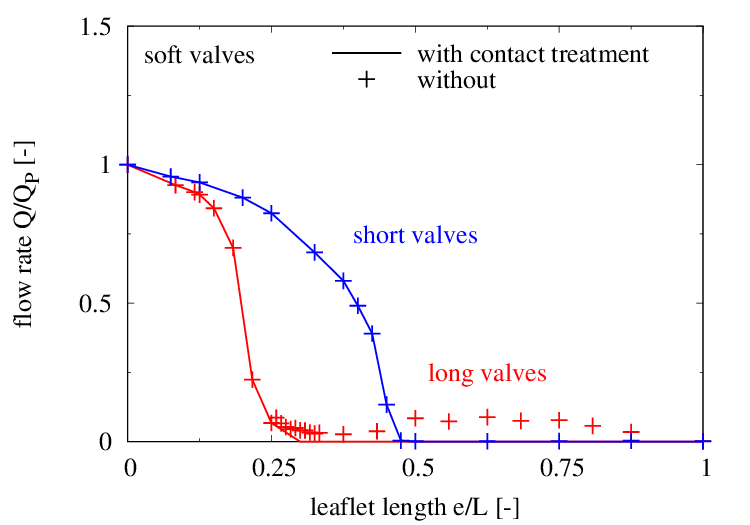}}
\quad
{\includegraphics[width=0.45\textwidth, angle=0]{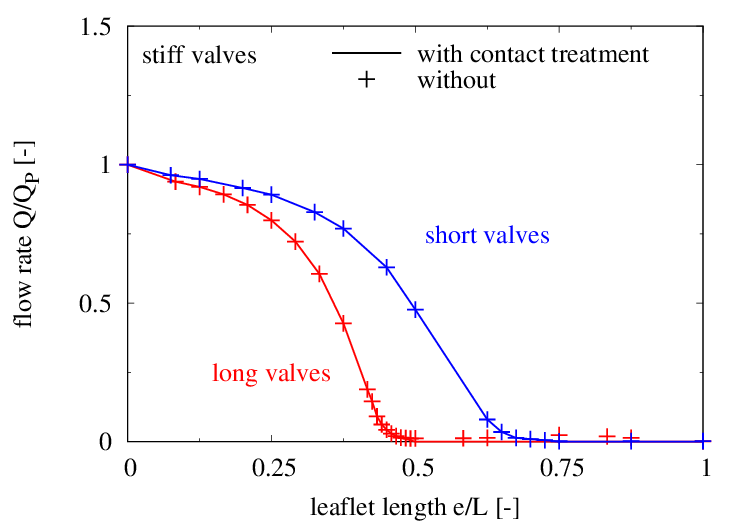}}
\caption{Comparison of computed reflux as a function of valve leaflet length for shorter and longer valves, with and without interleaflet contact treatment. The left panel shows results for soft valves, and the right panel for stiff valves. Contact treatment is essential for studying valves, as it enables complete closure of longer leaflets and consequently eliminates reflux.}
\label{fig:co_comp}
\end{figure}

For soft, longer valves with longer leaflets, expected to close and prevent reflux, a non-zero net flux is still observed, see the left panel of Fig.~\ref{fig:co_comp}.
This is due to fluid leakage through a gap that forms between the two leaflets. 
This gap is more pronounced in soft, longer valves, which have flexible leaflets that are displaced away from the vessel centerline by hydrodynamic forces, at low Reynolds numbers. 
In contrast, for stiff valves, the interleaflet gap tends to collapse, resulting in only negligible reflux (see the right panel of Fig.~\ref{fig:co_comp}).
To overcome this issue, both a contact treatment and a transition to bounce-back boundary conditions on the leaflets are implemented. 
Initially, the simulation runs with Fluid–Structure Interaction (FSI) and contact treatment until the valves reach a stable steady shape, in which the leaflets form a sealed coaptation zone. 
At this stage, FSI is switched off, and bounce-back conditions are used on the leaflets. 
Figure~\ref{fig:co_comp} shows the measured reflux with and without the contact treatment. 
The combination of contact treatment and transition to bounce-back boundary conditions on the leaflets results in zero net reflux for both stiff and soft valves, as expected.
\end{document}